%% file: main.tex
\documentclass[10pt,journal,compsoc]{IEEEtran}
\ifCLASSOPTIONcompsoc
  \usepackage[nocompress]{cite}
\else
  \usepackage{cite}
\fi
\ifCLASSINFOpdf
  \usepackage[pdftex]{graphicx}
\else
\fi
\usepackage{amsmath}
\usepackage{complexity}

\usepackage[ruled, vlined, linesnumbered]{algorithm2e}
\makeatletter
\newcommand{\removelatexerror}{\let\@latex@error\@gobble}
\makeatother
 \usepackage[caption=false,font=footnotesize,labelfont=sf,textfont=sf]{subfig}
\usepackage[hyphens,spaces,obeyspaces]{url}

\usepackage{color}
\usepackage{soul} 
\usepackage[backgroundcolor=yellow,textsize=tiny,textwidth=.9in]{todonotes}
\newcounter{todocounter}

\begin{document}
%
\title{Graph Drawing by Stochastic Gradient Descent}
%
%
%
%

\author{Jonathan~X.~Zheng,
        Samraat~Pawar,
		Dan~F.~M.~Goodman
\IEEEcompsocitemizethanks{
\IEEEcompsocthanksitem Jonathan Zheng is with Department of Electrical and Electronic Engineering, Imperial College London.\protect\\
E-mail: jxz12@ic.ac.uk
\IEEEcompsocthanksitem Samraat Pawar is with Department of Life Sciences, Imperial College London.
\IEEEcompsocthanksitem Dan Goodman is with Department of Electrical and Electronic Engineering, Imperial College London.
}
}

\IEEEtitleabstractindextext{%
\begin{abstract}
A popular method of
force-directed graph drawing
is multidimensional scaling using graph-theoretic distances as input.
We present an algorithm to minimize its
energy
function, known as stress, by using stochastic gradient descent (SGD) to move a single pair of vertices at a time. Our results show that SGD can reach lower stress levels faster and more consistently than majorization, 
without needing help from a good initialization.
We then show how the unique properties of SGD make it easier to produce constrained layouts than previous approaches.
We also show how SGD can be directly applied within the
sparse stress approximation
of Ortmann et al. \cite{ortmann2016sparse}, making the algorithm scalable up to large graphs.
\end{abstract}

}

\maketitle

\IEEEdisplaynontitleabstractindextext

%
\IEEEpeerreviewmaketitle

\IEEEraisesectionheading{\section{Introduction}\label{sec:introduction}}

%
%
%
%
\IEEEPARstart{G}{raphs} are a common data structure, used to describe everything from social networks to food webs, from metabolic pathways
to internet traffic. Any set of pairwise relationships between entities can be described by a graph, and the ever increasing amount of data being collected means that visualizing graphs for exploratory analysis has become an important task.

Node-link diagrams are an intuitive representation of graphs, where vertices are represented by dots, and edges by lines connecting them. A primary task is then to find suitable coordinates for these dots that represent the data faithfully.
However this is far from trivial, and the difficulty behind finding a good layout can be illustrated through a simple example. If we consider the problem of drawing a tetrahedron in 2D space, it is easy to see that no ideal layout exists where all edges have equal lengths. Even for
such a small graph with only four vertices,
there are too few dimensions available to provide sufficient degrees of freedom. The next logical question is: what layout gets as close as possible to this ideal?

Multidimensional scaling (MDS) is a technique to solve exactly this type of problem, that attempts to minimize the disparity between ideal and low-dimensional distances.
This is done by defining an equation to quantify the error in a layout, and then minimizing it. While this equation comes in many forms \cite{cox2000multidimensional},
\emph{distance scaling} is most commonly used for graphs \cite{brandes2008experimental}, where the error is defined as
%
\begin{equation}
\label{stress}
\text{stress}(\mathbf{X}) = \sum_{i<j} w_{ij}(||\mathbf{X}_i - \mathbf{X}_j|| - d_{ij})^2
\end{equation}
where $\mathbf{X}$ contains the coordinates of each vertex in low-dimensional space, and $d$ is the ideal distance between them. A weighting factor $w$ is used to either emphasize or dampen the importance of certain pairs. For the problem of graph layout, the most common approach is to set $d_{ij}$ to the shortest path distance between vertices $i$ and $j$, with $w_{ij} = d_{ij}^{-2}$ to offset the extra weight given to longer paths due to squaring the difference~\cite{brandes2008experimental}.

This definition was popularized for graph layout by Kamada and Kawai~\cite{kamada1989algorithm} who minimized the function using a localized 2D Newton-Raphson method, while within the MDS community Kruskal~\cite{kruskal1964multidimensional} originally used gradient descent~\cite{kruskal1964nonmetric}. This was later improved upon by De Leeuw~\cite{de1988convergence} with a method known as majorization,
which minimizes a complicated function by iteratively finding the 
true minima of a series of simpler functions, each of which touches the 
original function and is an upper bound for it~\cite{cox2000multidimensional}.
This was applied to graph layout by Gansner et al.~\cite{gansner2004graph} and
has been the state-of-the-art for the past decade.
For larger graphs, fully computing stress is not feasible, and so we review approximation methods in Section~\ref{sparse}.

This paper describes a method of minimizing stress by using stochastic gradient descent (SGD), which approximates the gradient of a sum of functions using the gradient of its individual terms. In our case this corresponds to moving a single pair of vertices at a time. The simplicity of each term in Equation~(\ref{stress}) also allows for some modifications to the step size which, combined with the added stochasticity, help to avoid local minima. We show the benefits of SGD over
majorization through experiment.


The structure of this paper is as follows: the algorithm is described and its subtleties are explored; an experimental study of its performance compared to majorization is presented; some real-world applications are shown that make use of the unique properties of SGD, including a method of making SGD scalable to large graphs by adapting the sparse approximation of Ortmann et al.~\cite{ortmann2016sparse}; and finally, we end with a discussion and ideas for future work.


\subsection{Constraint Relaxation}
The origin of our method is rooted in constrained graph layout, where a relaxation algorithm
has gained popularity due to its simplicity and versatility~\cite{dwyer2009scalable,bostock2011d3}.
It was first introduced in video game engines as a technique to quickly approximate the behavior of cloth, which is modeled as a planar mesh of vertices that maintains its edges at a fixed length.
A full physics simulation would represent each edge as a stiff spring, summing up and integrating over the resulting forces, but a realistic piece of cloth contains too many edges for this to be feasible.
To avoid this bottleneck, Jakobsen~\cite{jakobsen2001advanced} introduced the idea of considering each edge independently, moving a single pair of vertices at a time.
While this is a rather simple and perhaps naive idea, in practice the solution converges in very few iterations.

\begin{figure}
    \centering
    \includegraphics{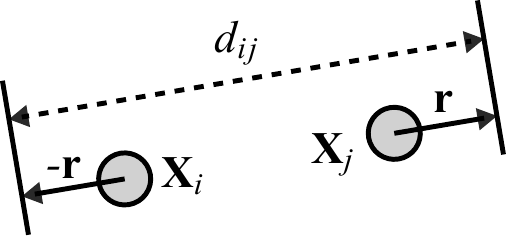}
    \caption{Satisfaction of the distance constraint described by Equation (\ref{constraint}).
    }
    \label{satisfaction}
\end{figure}

This was utilized by Dwyer~\cite{dwyer2009scalable}, who used the method in conjunction with a force-directed layout to achieve effects such as making edges point downwards, or fixing cycles around the edge of a wheel. To define it properly in the case of maintaining a distance $d_{ij}$ between two vertices $X_i$ and $X_j$, this movement, known henceforth as a constraint, can be written as
\begin{equation}
\label{constraint}
||\mathbf{X}_i - \mathbf{X}_j|| \leftarrow d_{ij}
\end{equation}
and is satisfied by moving $\mathbf{X}_i$ and $\mathbf{X}_j$ in opposite directions by a vector
\begin{equation}
\mathbf{r}
= \frac{||\mathbf{X}_i - \mathbf{X}_j||-d_{ij}}{2}\frac{\mathbf{X}_i - \mathbf{X}_j}{||\mathbf{X}_i - \mathbf{X}_j||}.
\end{equation}
This can be seen as a diagram in Figure~\ref{satisfaction}, and is analogous to decompressing an infinitely stiff spring of length $d_{ij}$.
%

Rewriting Equation~(\ref{stress}) as
\begin{gather}
\label{stress-terms}
\text{stress}(\mathbf{X}) = \sum_{i<j} Q_{ij}(\mathbf{X}),\\
\label{qij}
Q_{ij}(\mathbf{X}) = w_{ij}(||\mathbf{X}_i - \mathbf{X}_j|| - d_{ij})^2,
\end{gather}
it can be seen that if every term $Q_{ij}$ in the summation is satisfied as a constraint~(\ref{constraint}), then the total stress is zero, corresponding to an ideal layout. 
This is exactly the idea behind our method---we
replace the force-directed
component by instead placing a constraint on every possible pair of vertices, satisfying them one by one as above. However zero stress
is almost always impossible, for the same reasons that the aforementioned tetrahedron cannot be embedded in 2D. In such situations, simply satisfying constraints as
above
does not lead to convergence, but we will now describe an extension that does.

\section{Stochastic Gradient Descent}
\label{wcr description}
Our modifications to the algorithm described above can be understood by first noticing that satisfying a constraint is equivalent to moving both vertices in the direction of the gradient of a stress term $Q_{ij}$
\begin{equation}
  \label{gradient}
\frac{\partial Q_{ij}}{\partial\mathbf{X}_i}=\frac{\partial}{\partial\mathbf{X}_i}w_{ij}(||\mathbf{X}_i - \mathbf{X}_j|| - d_{ij})^2
= 4w_{ij}\mathbf{r}.
\end{equation}
%
We can compute the full gradient $\partial Q_{ij}/\partial\mathbf{X}$ as
\begin{equation}
\frac{\partial Q_{ij}}{\partial\mathbf{X}_k}=\begin{cases}
4w_{ij}\mathbf{r} & \mbox{if $k=i$} \\
-4w_{ij}\mathbf{r} & \mbox{if $k=j$} \\
0 & \mbox{otherwise.} \\
\end{cases}
\end{equation}
Directly applying stochastic gradient descent to minimize
stress
would involve repeatedly randomly selecting a term $Q_{ij}$ and applying the iterative formula $\mathbf{X}\leftarrow \mathbf{X}-\eta\nabla Q_{ij}(\mathbf{X})$, where $\eta$ is a step size that tends towards 0 as the iteration number increases.
Note that since the gradient is zero with respect to all $\mathbf{X}_k$ other than $\mathbf{X}_i$ and $\mathbf{X}_j$, it suffices to update the positions of $\mathbf{X}_i$ and $\mathbf{X}_j$ by
\begin{equation}
\label{SGD step}
\begin{bmatrix}\mathbf{X}_i\\\mathbf{X}_j\end{bmatrix}
\leftarrow
\begin{bmatrix}\mathbf{X}_i\\\mathbf{X}_j\end{bmatrix}+
\begin{bmatrix}\Delta \mathbf{X}_i\\\Delta \mathbf{X}_j\end{bmatrix}
=
\begin{bmatrix}\mathbf{X}_i\\\mathbf{X}_j\end{bmatrix}-
4w_{ij}\eta
\begin{bmatrix}\mathbf{r}\\-\mathbf{r}\end{bmatrix}.
\end{equation}

\begin{figure}
    \centering
    \includegraphics[width=\linewidth]{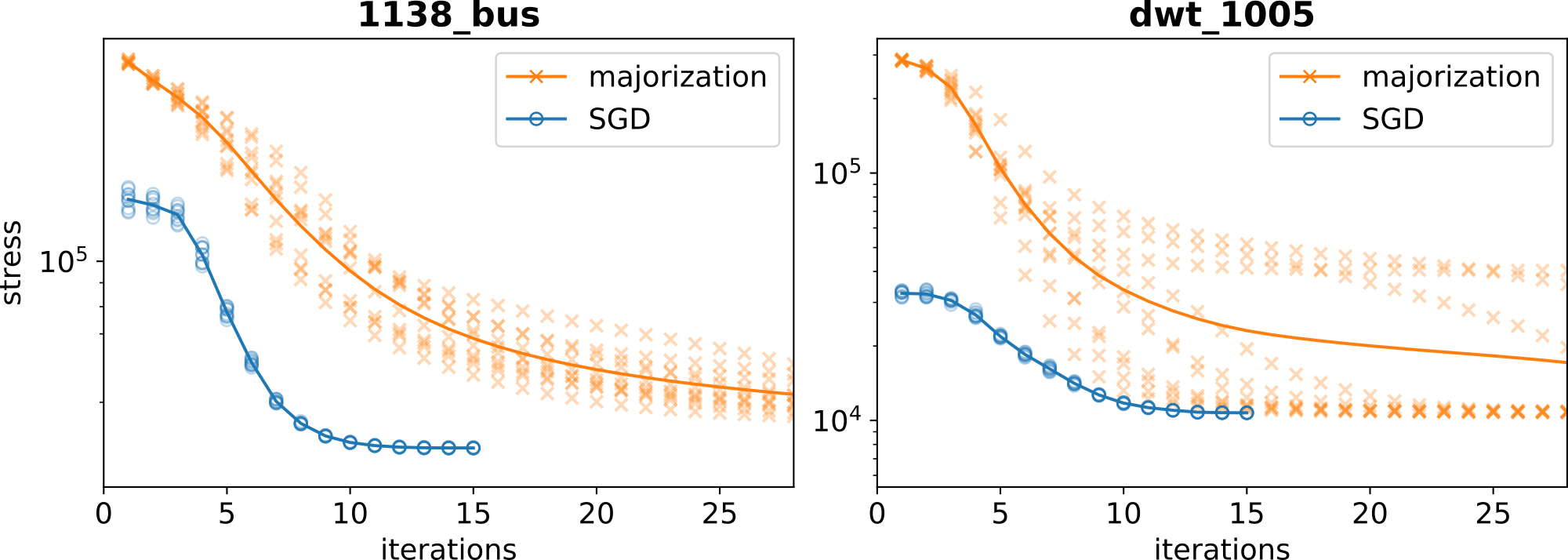}
    \caption{Plots of stress for SGD and majorization on the graphs \texttt{1138\_bus} and \texttt{dwt\_1005}, each initialized randomly within a 1$\times$1 square.
    The circles and crosses show stress on each iteration over 10 runs, with the line running through the mean.
    Initial stress values are omitted.
    SGD is clearly more consistent, always reaching lower stress levels than majorization ever manages in hundreds of iterations on \texttt{1138\_bus}.
    They both reach the same overall minimum on the more mesh-like \texttt{dwt\_1005}, but majorization often gets stuck on a particularly dangerous local minimum, shown by its diverging paths.
    A more detailed timing analysis on a wide variety of other examples can be seen in Section~\ref{results}.
	}
    \label{stress plots}
\end{figure}

The constraint relaxation of the previous section is therefore equivalent to a special case of SGD where $w_{ij}=1$ and $\eta=1/4$. Writing $\mu=4w_{ij}\eta$ as the coefficient of $\mathbf{r}$ we can see that $Q_{ij}\leftarrow 0$ when $\mu=1$ and decreases monotonically from $\mu=0$ to $\mu=1$. Since we have this extra geometric structure that is not normally available in SGD settings, we investigated a modified SGD algorithm in which we set a hard upper limit of $\mu\leq 1$:
\begin{equation}
\label{mu}
\begin{aligned}
\Delta\mathbf{X}_i &= -\Delta\mathbf{X}_j = -\mu\, \mathbf{r},\\
\mu&=\min\{\,w_{ij}\eta, \, 1\,\}.
\end{aligned}
\end{equation}
This modified algorithm makes updates that are identical to standard SGD when $\eta$ is sufficiently small,
\begin{equation}
\label{eta-sufficiently-small}
\eta<\frac{1}{w_{\max}}.
\end{equation}
Since this will always eventually be the case, it has the same asymptotic convergence properties as standard SGD, which we discuss in Section~\ref{unlimited iterations}.
However, we find that introducing this upper limit on $\mu$ allows for much larger initial step sizes than standard SGD,
yielding much faster convergence without getting stuck in local minima. We show by experiment that this is true for a wide range of graphs (except for a single specific case, see Section~\ref{quality}).
In addition, we use random reshuffling of terms unless otherwise stated (see Section~\ref{randomization}). We define a full pass through all the terms $Q_{ij}$ as a single \emph{iteration}, while a single application of Equation~(\ref{SGD step}) will be known as as a \emph{step}.
From now on, we will refer to our modified SGD algorithm simply as SGD.

Plots of stress achieved using SGD compared to majorization are presented briefly in Figure~\ref{stress plots}, and in more detail in Section~\ref{results}. Pseudocode is shown in Algorithm~\ref{pseudocode} (Figure~\ref{algorithm1}).
All results in this paper have vertex positions initialized uniformly randomly within a 1$\times$1 square.
Unless stated otherwise, graph data is from the SuiteSparse Matrix Collection~\cite{davis2011university}. Tests were performed using C\# running in Visual Studio, on an Intel Core i7-4790 CPU with 16GB of RAM.

\begin{figure}
\removelatexerror

\begin{algorithm}[H]
\SetKwProg{SGD}{SGD}{}{}
\SetKwInOut{Input}{inputs}
\SetKwInOut{Output}{output}
\SetKwFor{ForEach}{foreach}{:}{}
\SetKwFor{For}{for}{:}{}
\SetKwIF{If}{ElseIf}{Else}{if}{:}{elif}{else}{}

\SGD{\emph{\textbf{(}}$G$\emph{\textbf{):}}}{
    \Input{graph $G=(V,E)$}
    \Output{$k$-dimensional layout $\mathbf{X}$ with $n$ vertices}
    
    $d_{\{i,j\}} \leftarrow ShortestPaths(G)$
    \label{code:bacon}
    
    
    $\mathbf{X} \leftarrow RandomMatrix(n,k)$
    \label{code:init}

    
    \For{$\eta$ in annealing schedule}{
    \label{code:annealing}
    
    
        \ForEach{$\{i,j:i<j\}$ in random order}{

            $\mu \leftarrow w_{ij} \eta$
            
            \If{$\mu > 1$}{ \label{code:if1}
                $\mu \leftarrow 1$ \label{code:if2}
            }
            
            $\mathbf{r} \leftarrow \frac{||\mathbf{X}_i - \mathbf{X}_j||-d_{ij}}{2}\frac{\mathbf{X}_i - \mathbf{X}_j}{||\mathbf{X}_i - \mathbf{X}_j||}$
            
            $\mathbf{X}_i \leftarrow \mathbf{X}_i - \mu\,\mathbf{r}$
            \label{code:satisfaction1}
            
            $\mathbf{X}_j \leftarrow \mathbf{X}_j + \mu\,\mathbf{r}$
            \label{code:satisfaction2}
        }
    }
}
\caption{Stochastic Gradient Descent}
\label{pseudocode}
\end{algorithm}

\caption{
Pseudocode for the algorithm described in Section~\ref{wcr description}.
The results in this paper initialize positions randomly within a 1$\times$1 square on line~\ref{code:init}.
The annealing schedule on line~\ref{code:annealing} is explained in Section~\ref{cooling}.
}
    \label{algorithm1}
\end{figure}

\subsection{Step Size Annealing}
\label{cooling}
Choosing a good step size $\eta$ is crucial to the performance of SGD~\cite{darken1992learning}, and a typical implementation can involve complex algorithms for tuning the step size to the problem at hand~\cite{ruder2016overview}.
Most of these methods do not apply here for two reasons. First, due to the limit on the step size in Equation~(\ref{mu}), we can and do use much larger step sizes than standard SGD would allow. Second, many of these methods use previous gradients to inform the step size; we only update the positions of the two vertices directly involved, so storing and applying previous gradients is inefficient to the point of increasing the asymptotic complexity of the algorithm.


Even ignoring such adaptive methods, the full space of possible
annealing schedules is too large to investigate in full, and the results can differ depending on the graph. We therefore investigated a limited subset of possible schedules, taking the mean final stress across a wide range of graphs as the performance criterion
(the full set of graphs considered in Section~\ref{results}).
We consider two use cases: one where time is a limiting factor and so the number of iterations is predetermined, and another where the algorithm may continue until the layout has converged to within a desired accuracy.

\subsubsection{Fixed Number of Iterations}
\label{fixediterations}
We consider a step size that starts at a maximum value $\eta=\eta_\mathrm{max}$ at the first iteration $t=0$, and decreases monotonically to $\eta=\eta_\mathrm{min}$ at the final iteration $t=t_\mathrm{max}-1$.
Large values of $\eta$ result in all $\mu$ capped at 1, and very small values will result in little to no movement of vertices. Because we wish to work within the useful range in between these extremes, we set
\begin{equation}
\label{etamaxmin}
\eta_{\max} = \frac{1}{w_{\min}} \,,\;\; \eta_{\min} = \frac{\varepsilon}{w_{\max}}.
\end{equation}
In our case $w_{ij} = d_{ij}^{-2}$ so $w_{\min}$ is inversely proportional to the diameter of the graph $d_{\max}$, and $w_{\max}$ to the smallest edge length $d_{\min}$.
This choice of $\eta_{\max}$ ensures that all $\mu = 1$ for the first iteration, which appears to be desirable in order to avoid local minima, while the choice of $\eta_{\min}$ ensures that even the strongest constraints reach a small value of $\mu = \varepsilon$ for the final iteration.

\begin{figure}
    \centering
    \includegraphics[width=\linewidth]{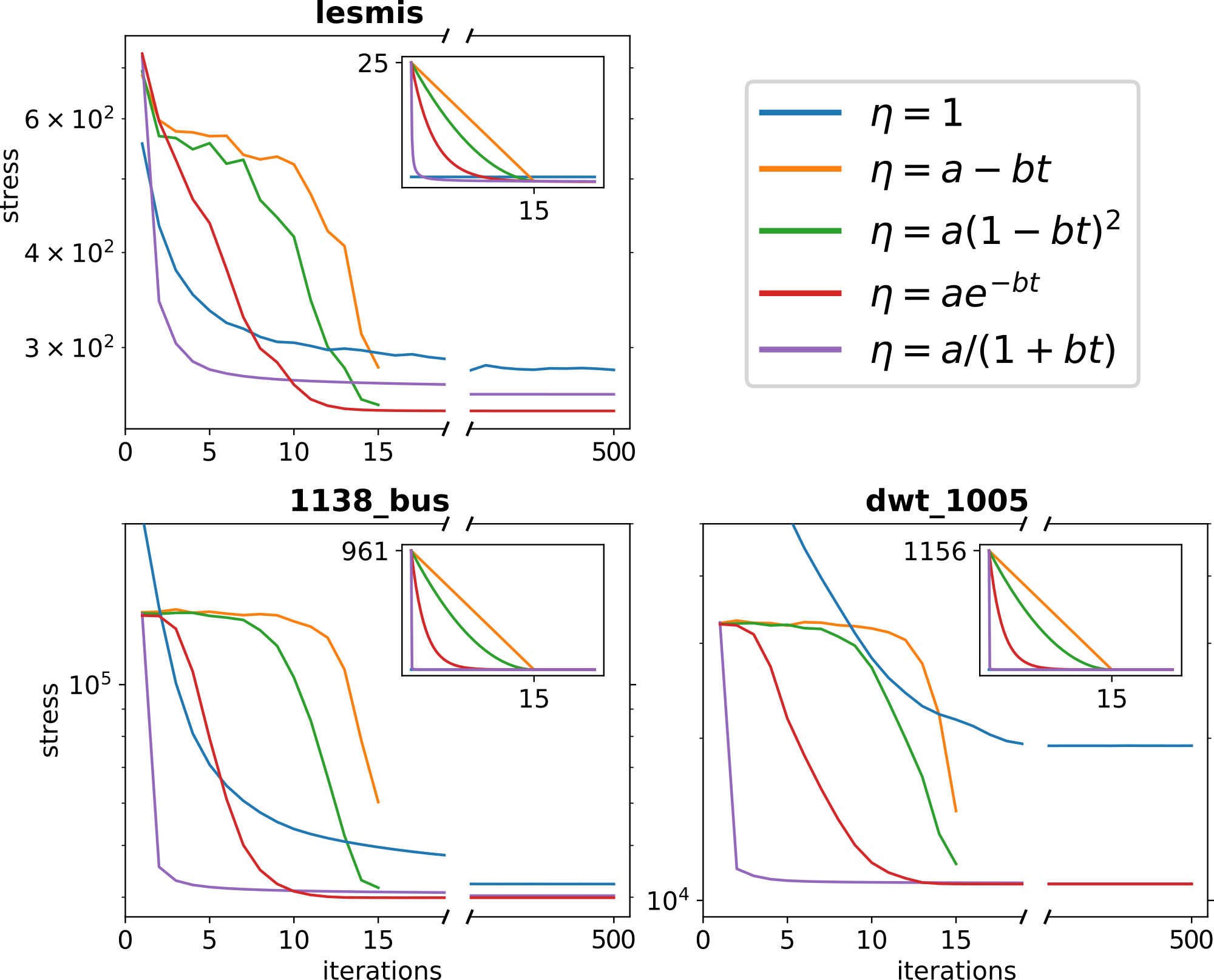}
    \caption{Plots of mean stress against iterations over 25 runs for the annealing schedules discussed in Section~\ref{cooling}, with $t_{\max}=15$ and $\varepsilon=0.1$ in all cases. The exact schedules used are shown in the top right of every plot.
    To approximate behavior given unlimited time, schedules were run for 500 iterations.
    \texttt{1138\_bus} shows the typical behavior of $\eta= ae^{-bt}$ reaching lower stress and $\eta= a/(1+bt)$ never quite catching up;
    \texttt{lesmis} shows that this applies to smaller graphs as well;
    \texttt{dwt\_1005} emphasizes the importance of larger step sizes, as the constant $\eta=1$ struggles to jump over large local minima.
    Note that on this easier graph $\eta=a/(1+bt)$ is enough to reach good minima in very few iterations.
    }
    \label{annealing}
\end{figure}

\begin{figure}
    \centering
    \includegraphics[width=\linewidth]{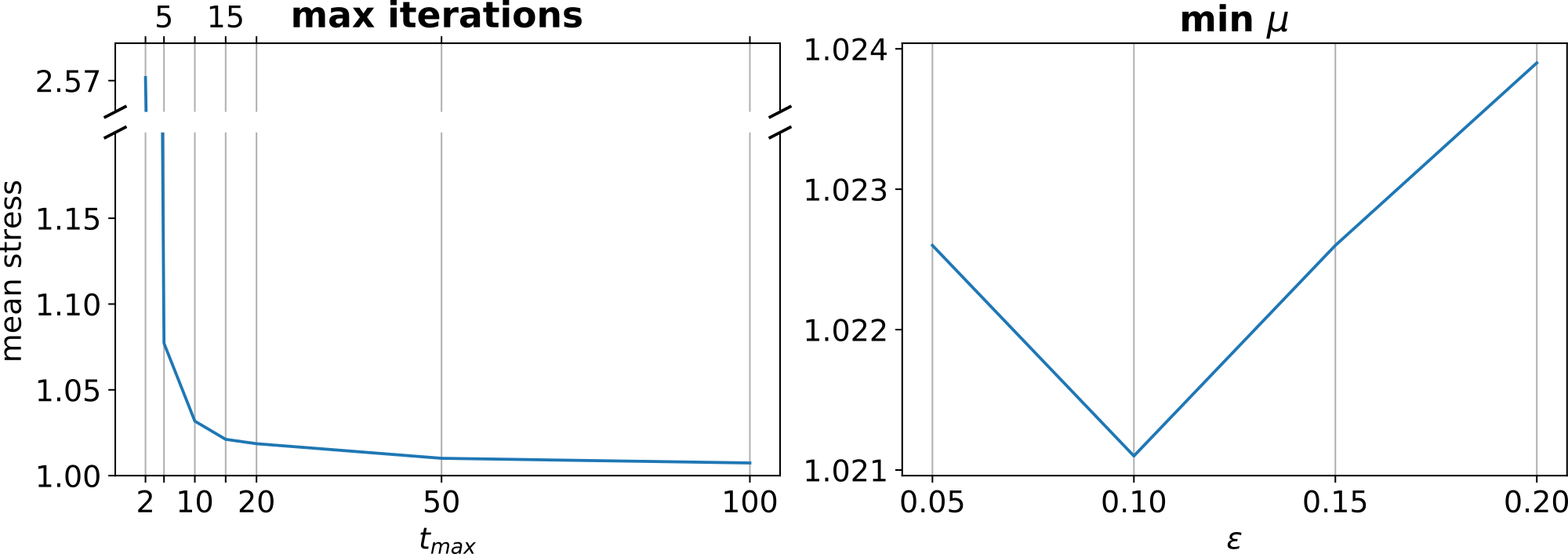}
    \caption{Plots of mean stress over 25 runs on all graphs in Section~\ref{results} when varying the parameters $t_{\max}$ or $\varepsilon$ on Equation~(\ref{expodecay}), normalized to the best values from Figure~\ref{systematic}.
    There are clear diminishing returns when increasing $t_{\max}$, so we chose $t_{\max}=15$ as a trade-off between speed and quality.
    $\varepsilon=0.1$ is close to optimum for this value.
    }
    \label{parameters}
\end{figure}

We computed the performance for various schedules $\eta(t)$ where $t$ is the iteration number, constrained to $\eta(0)=\eta_{\max}$ and $\eta(t_\mathrm{max}-1)=\eta_{\min}$ (except for the special case $\eta(t)=1$).
In each panel of Figure~\ref{annealing} we vary the form of the function $\eta(t)$ for a fixed choice of $t_\mathrm{max}$ and $\varepsilon$. The best form of $\eta(t)$ appears to be the exponential decay given by the equation
\begin{equation} \label{expodecay}
\eta_1(t) = \eta_{\max} e^{-\lambda t}.
\end{equation}
In addition we varied the parameters $t_\mathrm{max}$ and $\varepsilon$ for this $\eta(t)$ (see Figure~\ref{parameters}).
Increasing $t_\mathrm{max}$ always improves the quality but also increases computation time, so we chose $t_\mathrm{max}=15$ as a reasonable compromise between speed and quality.
The choice $\varepsilon=0.1$ appears to be close to optimal for this $t_{\max}$.
With this number of iterations most of the gains had already been made and further ones gave diminishing returns, although for particular applications another choice may be more appropriate.

It is common in SGD to use a schedule
$\eta = \Theta(1/t)$~\cite{darken1992learning},
however for the small number of iterations considered here, the large initial step sizes cause $\eta$ to decay too quickly in the beginning, leading to worse local minima. 
Exponential decay drops faster than $1/t$ as $t\rightarrow\infty$, but $1/t$ drops faster in early iterations given fixed values at $\eta(0)$ and $\eta(t_{\max}-1)$, as shown by the inset panels in Figure~\ref{annealing}.

\subsubsection{Unlimited Iterations/Convergence}
\label{unlimited iterations}
The schedule described above works well in practice for a fixed number of iterations, but given more time it can be desirable to let the algorithm run for longer to produce an optimal layout.
Here we describe a schedule that is guaranteed to converge, and a stopping criterion to prevent the algorithm from wasting iterations on negligible movements.

A proof of convergence for SGD is well known in the machine learning literature~\cite{bottou2012stochastic}, and requires an annealing schedule that satisfies
\begin{equation}
\label{convergence}
\sum_{t=0}^{\infty}\eta(t) = \infty
\quad \text{and} \quad
\sum_{t=0}^{\infty}\eta(t)^2 < \infty.
\end{equation}
This is guaranteed to reach global minima under conditions slightly weaker than convexity~\cite{bottou1998online}.
Intuitively, the first summation ensures the decay is slow enough to reach the minimum no matter how far away we initialize, and the second ensures fast enough decay to converge to, rather than bounce around the minimum~\cite{welling2011bayesian}.
In the context of non-convex functions, like the stress equation considered in this paper, such a proof only holds for convergence to a stationary point that may be a saddle~\cite{bottou1998online}.
There is also recent work proving convergence to local minima in specific classes of non-convex functions~\cite{ge2015escaping}.



Since we cannot guarantee a global minimum with any choice of schedule, the best we can do is to choose a schedule that will converge to a stationary point. A commonly used schedule that guarantees this is $\eta=\Theta(1/t)$, as this satisfies Equation~(\ref{convergence}). However, in the previous section we noted that this schedule gets stuck in poor local minima. We therefore use a mixed schedule. When $t$ is small, $\eta(t)=\eta_1(t)$ follows the exponential schedule of the previous section, because in practice this avoids poor local minima. When $t$ is large we then switch to a $1/t$ schedule to guarantee convergence to a stationary point:
\begin{equation}
\eta_{2}(t + \tau) =
\frac{w_{\max}^{-1}}{1+\lambda t}
\quad \text{when}
\quad t>\tau\: :\:
\eta_{1}(\tau) = w_{\max}^{-1}.
\end{equation}
%
The cross-over value $\tau$ is the iteration at which the limit in Equation~(\ref{mu}) stops capping $\mu$ and our algorithm becomes standard SGD.
Since we have more iterations to work with, we also choose $t_{\max}=30$ in order to further improve avoidance of local minima. This choice is sufficient to give even or better mean performance than majorization after convergence across every graph we tested except for one (see Section~\ref{quality}), but again depending on the application another choice may be more suitable.

Finally, we introduce a suitable stopping criterion. Since SGD does not guarantee the monotonic decrease of stress~\cite{darken1992learning}, we cannot use the majorization heuristic adopted by Gansner et al.~\cite{gansner2004graph}, which stops when the relative change in stress drops below a certain threshold.
However we can guarantee that each time a constraint is satisfied, its corresponding term within the summation does decrease.
We therefore estimate how close we are to convergence by tracking the maximum distance a vertex is moved by a single step over the previous iteration, and stop when this crosses a threshold
\begin{equation}
\max||\Delta\mathbf{X}|| < \delta.
\end{equation}
%
We find that a value of $\delta=0.03$ works well in practice.

Thus we have designed two schedules: one for a fixed number of iterations, and one that continues until convergence.
Results using both of these are presented in Section~\ref{results}.
It is important to note that these schedules use simple heuristics, and the exact nature of the data will affect the results.
However we find that they are robust across a wide variety of graphs, as all the results shown in this paper use these two schedules.

\subsection{Randomization}
\label{randomization}

\begin{figure}
    \centering
    \includegraphics[width=\linewidth]{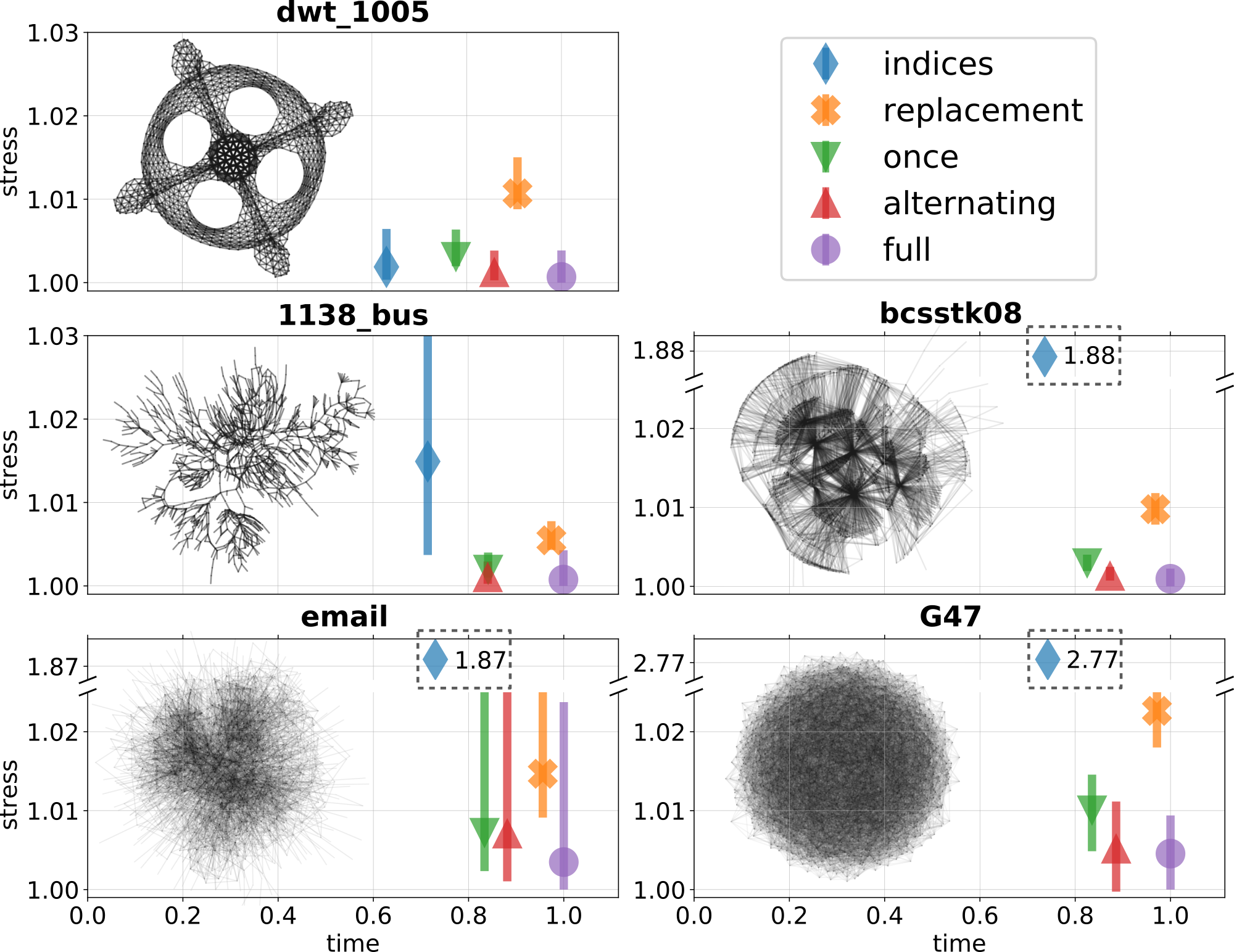}
    \caption{Stress against time taken using different degrees of randomization, over 50 runs from different random starting configurations. Markers indicate mean stress, with vertical bars ranging from best to worst over all runs.
    Both stress and time are normalized to the absolute minimum and maximum respectively over any run
    }
    \label{random}
\end{figure}

An important consideration is the order in which constraints are satisfied, as naive iteration can introduce biases that cause the algorithm to get caught in local minima.
The original method behind SGD proposed by Robbins and Monro~\cite{robbins1951stochastic} randomizes with replacement, meaning that a random term is picked every time with no guarantee as to how often a term will be picked. Some variants perform random reshuffling (RR) which guarantees that every term is processed once on each iteration. Under certain conditions it can be proven analytically that RR converges faster~\cite{gurbuzbalaban2015random}, and our results support this.

Unfortunately adding randomness incurs a penalty in speed, due to the cost of both random number generation and reduced data cache prefetching.
We found that this overhead is non-trivial, with iterations taking up to 60\% longer with random reshuffling compared to looping in order. 
We explored the trade-offs between more randomness for better convergence but slower iterations, versus less randomness for slower convergence but faster iterations.
We tried five different degrees of randomness: shuffling only the indices themselves, which removes any bias inherent to the data but still makes use of the cache by iterating in order; randomizing with replacement; shuffling the order of terms once; shuffling twice and alternating between the two orders; and shuffling on every iteration.
The results can be seen in Figure~\ref{random}.

We selected five different graphs, each with around 1000 vertices, and show a corresponding good layout for each to visualize the differences between them.
More mesh-like graphs such as \texttt{dwt\_1005} do not benefit much from added randomness, and receive large gains in speed for a small hit to quality.
As graphs get more difficult to draw, shuffling only indices quickly becomes ineffective,
with mean stress levels off by orders of magnitude on the plots with broken axes.
The graph \texttt{email} is a social network, which tend to be very difficult to draw as their global minima are difficult to find. The drop in quality when reducing the randomness reflects this. \texttt{G47} is a random graph and has the highest stress, but is easier to draw since there are many minima close to global that are all relatively easy to find.

Although RR is the most expensive method, it is only slightly more expensive and consistently performs best. However if speed is the most important concern, alternating between two random shuffles gives stress levels that are in many cases almost as good, at a slightly reduced cost.
We use RR for the rest of the results here.

\section{Experimental Comparison}
\label{results}

\begin{figure*}
    \centering
    \includegraphics[width=\textwidth]{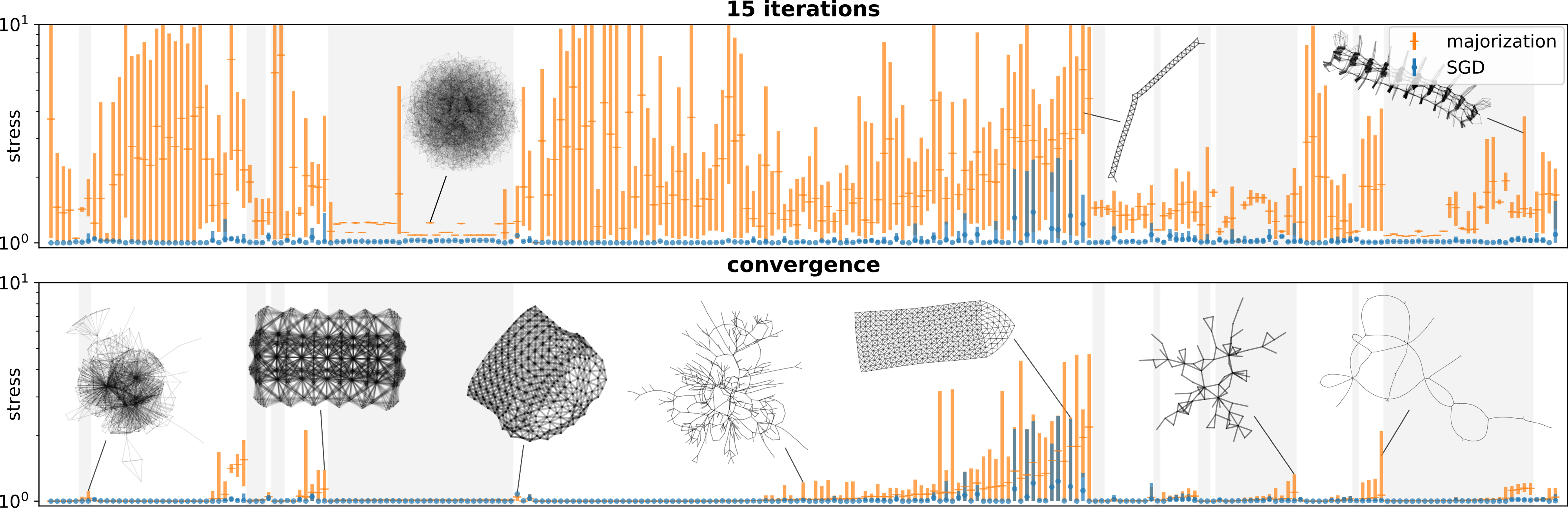}
    \caption{Stress achieved over 25 runs on every symmetric matrix with 1000 or fewer vertices in the SparseSuite Matrix Collection~\cite{davis2011university}, comprising 243 graphs in total. Markers indicate the mean over all runs, with bars ranging from minimum to maximum on any run.
    The top plot shows values reached after 15 iterations, and the bottom after convergence.
    Stress values are normalized to the lowest value achieved on all runs for either algorithm, as a baseline `correct' layout.
    Each graph in the collection is also assigned to a group as metadata, and graphs that share a group tend to have similar topologies; we keep groups together, ordered alphabetically, and within each group sort by difference in mean stress after convergence. Thus both plots have the same order. Groups are demarcated by the alternating shaded background. Layouts shown are, from left to right, top then bottom:
\texttt{G15}, 
\texttt{dwt\_66}, 
\texttt{orbitRaising\_2}, 
\texttt{celegans\_metabolic},
\texttt{ex2},
\texttt{dwt\_307},
\texttt{494\_bus},
\texttt{dwt\_361},
\texttt{Sandi\_authors},
\texttt{S10PI\_n1}.
	An animated version of the top plot is included in the supplemental material.
    }
    \label{systematic}
\end{figure*}

To test the effectiveness of our algorithm, we follow Khoury et al.~\cite{khoury2012drawing} and use symmetric sparse matrices from the SuiteSparse Matrix Collection~\cite{davis2011university} as a benchmark.
We ran both SGD and majorization on every graph with 1000 or fewer vertices, and compared the range of stress levels reached after 15 iterations and until convergence, using the two schedules described Section~\ref{cooling}. These results can be seen in Figure~\ref{systematic}.
We also chose a representative selection of larger graphs for more detailed timing results, showing multiple implementations of majorization and the time course of convergence, which can be seen in Figure~\ref{appendix}.

\begin{figure}
    \centering
    \includegraphics[width=\linewidth]{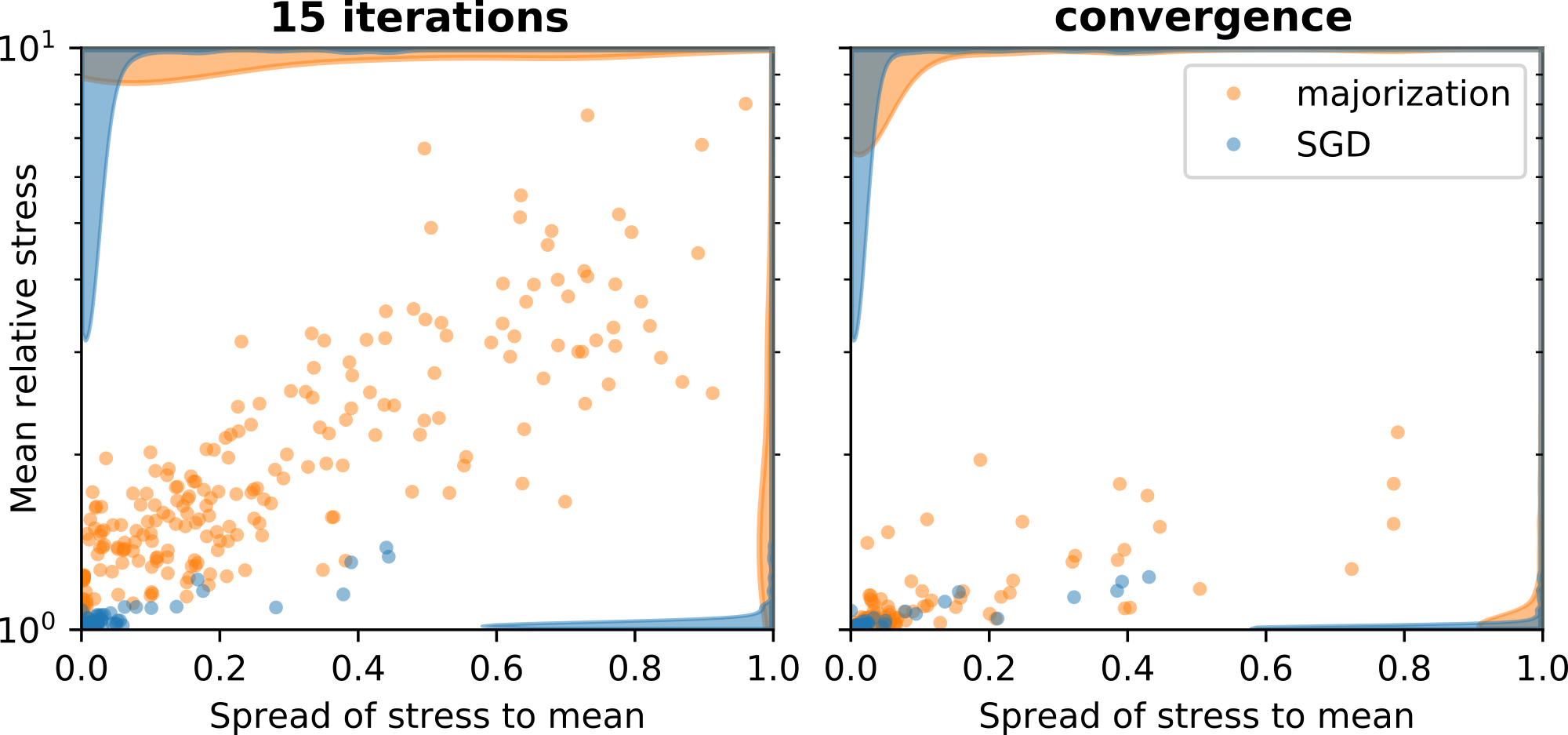}
    \caption{Scatter plots of mean stress relative to the best achieved, against the spread of the stress, measured as the coefficient of variation (standard deviation over mean), using the same results as Figure~\ref{systematic}.
The shaded regions on the top and right of each panel show the density of the mean (right) and spread (top) of the stress, computed using the \texttt{gaussian\_kde} function of SciPy~\cite{scipy}.
    }
    \label{summary}
\end{figure}

\subsection{Quality}
\label{quality}
We can see from Figure~\ref{systematic} that SGD reaches the same low stress levels on almost every run. While majorization is proven to monotonically decrease stress \cite{gansner2004graph}, 
it can often struggle with local minima. This can be clearly seen in Figure~\ref{summary}, as majorization consistently shows larger variance in its stress trajectories from different starting configurations.

The layouts displayed in Figure~\ref{systematic} were chosen to highlight the effects of different types of graphs. From left to right, top then bottom: \texttt{G15} is a random graph with nodes decreasing in mean degree. These random graphs reach consistent stress levels with both algorithms, as their lack of structure results in many minima close to global.
\texttt{dwt\_66} is an example of a graph that majorization struggles with, as it is very long and often has multiple twists that majorization cannot unravel.
\texttt{orbitRaising\_2} is a similar example, but in this case majorization also never reaches the global minimum, even after convergence.
\texttt{celegans\_metabolic} is a metabolic pathway that is around as densely packed as a graph worth drawing gets.
SGD consistently outperforms majorization here too.
Many of the largest ranges in the plot are from graphs similar to \texttt{ex2}; grids are difficult to fully unfold, and majorization often struggles with their many local minima.

\begin{figure}
    \centering
    \includegraphics[width=\linewidth]{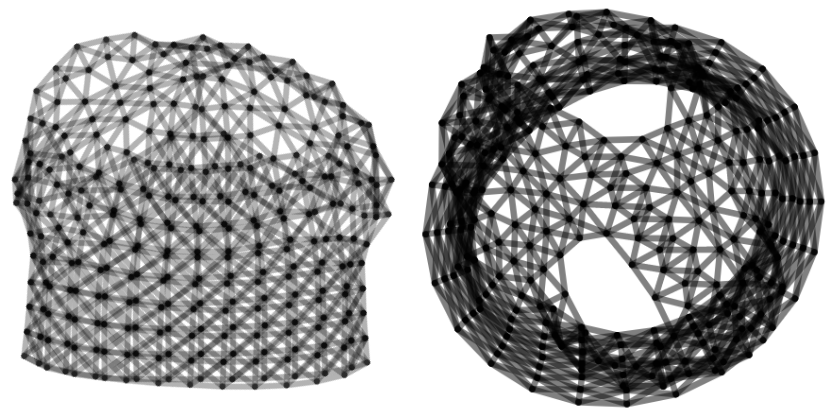}
    \caption{Layouts from \texttt{dwt\_307}, the only one of the 243 graphs considered where majorization (left) yields lower stress than SGD (right).
	}
    \label{cylinder}
\end{figure}

On the other hand, \texttt{dwt\_307} is the one graph (of the 243 we investigated) where majorization reaches lower stress than SGD as a result of our modifications to standard SGD (Figure~\ref{cylinder}).

\texttt{494\_bus} is an example of the type of graph where Equation~(\ref{stress}) produces better layouts than other popular models. Its symmetry is clear here, whereas other force-directed algorithms can fail to show this due to the \textit{peripheral effect}~\cite{hu2005efficient}.
\texttt{dwt\_361} is an example of the type of graph that both SGD and majorization struggle with: long graphs that can twist. A twist in a graph constitutes a deep local minimum that iterative methods struggle with in general, and SGD is still susceptible to this issue.
\texttt{Sandi\_authors} is a small graph, but with some densely packed sections that can become stuck behind each other, something that majorization often struggles with.
And finally, \texttt{S10PI\_n1} is a long graph that does not get twisted and so SGD deals with it perfectly well, but its long strands still tend to give majorization problems.



\subsection{Speed}

Our results show that SGD converges to low stress levels in far fewer iterations than majorization. Graphs are laid out in only 15 iterations in the top plot in Figure~\ref{systematic}, and there is not much improvement to be gained from using the convergent schedule to let the algorithm run for longer. This indicates that most global minima can be found in very few iterations, making SGD especially suited for real-time applications such as interactive layout.
Our stopping criterion for majorization was for relative decrease in stress to be less than $10^{-5}$, which is ten times more forgiving than originally suggested by Gansner et al.~\cite{gansner2004graph}, as we found $10^{-4}$ was not lenient enough to be confident that it had settled completely. Given enough time, majorization does find good minima more often than not, but can still settle in local minima and in some cases never finds the best configuration regardless of initialization.
Majorization also takes many more iterations to converge than SGD, with means of 237 and 106 iterations respectively.

\begin{figure*}
    \centering
    \includegraphics[width=\textwidth]{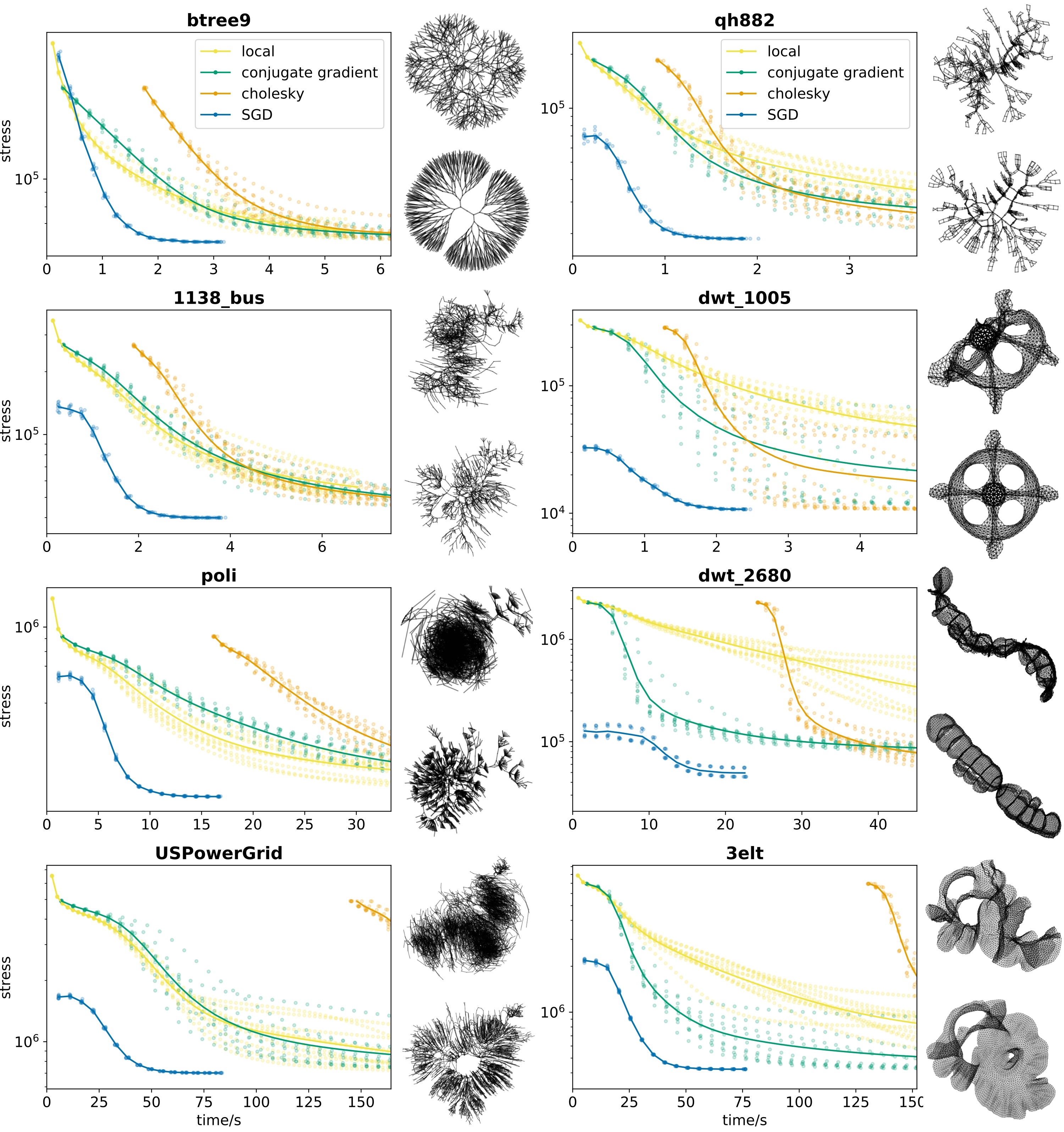}
    \caption{Graphs of stress against time for all three implementations of majorization, and SGD using the 15 iteration annealing schedule. 10 runs were used for each graph, and the line plots run through the mean stress and time per iteration.
    Graphs were considered as unweighted
    and layouts were initialized randomly within a 1$\times$1 square.
    The graph \texttt{btree9} is a binary tree of depth 9. 
    Initial stress values are omitted, which is the cause of the horizontal offset to Cholesky due to its longer initialization time.
    Note that if we remove the time for the first iteration Cholesky outperforms the other implementations of majorization, but still does not ever drop below the curve for SGD.
    The layouts show examples of what the graphs look like after 15 iterations, with Cholesky on top and SGD on bottom.
    }
    \label{appendix}
\end{figure*}

The real-world time per iteration must also be considered, even though both share a complexity $O(n^2)$. We adapted the Cholesky factorization routine from Numerical Recipes~\cite{press1996numerical} to C\#, and found that iterations are around 40\% faster than SGD. However the initial decomposition before back-substitution requires $n^3/6$ iterations involving a multiply and a subtract~\cite{press1996numerical}, so the total time quickly tips in favor of SGD.
Conjugate gradient (CG), with tolerance 0.1 and max iterations 10 as in~\cite{gansner2013maxent}, is an iterative method itself to solve the majorizing function and so iterates slower than Cholesky and SGD, but often beats out Cholesky overall when fewer iterations are necessary.
CG and Cholesky also both benefit from optimized matrix multiplication routines~\cite{gansner2004graph} that we did not try here.
Localized majorization, which is used to majorize the sparse model in Section~\ref{sparse}, iterates fastest of all but converges slower.
It is also worth noting that over-shooting has been used before in the context of majorization to achieve an average of 1.5 times speedup~\cite{wang2012fast}.
Plots of stress against real time can be seen in Figure~\ref{appendix}.

\section{Applications}
Some of the properties of SGD, in particular the fact that each edge is considered separately along with the ability to consistently avoid local minima well, make SGD well suited to variants such as constrained layout. We will now describe some recipes for examples of this, each applied to various real-world graphs in order to show the merits of their use.
Note that these applications are also possible with majorization, but can require more drastic modifications in order to apply them successfully.

\subsection{Focusing on a Vertex}
\label{focus}

\begin{figure}
    \centering
    \includegraphics[height=0.85\linewidth]{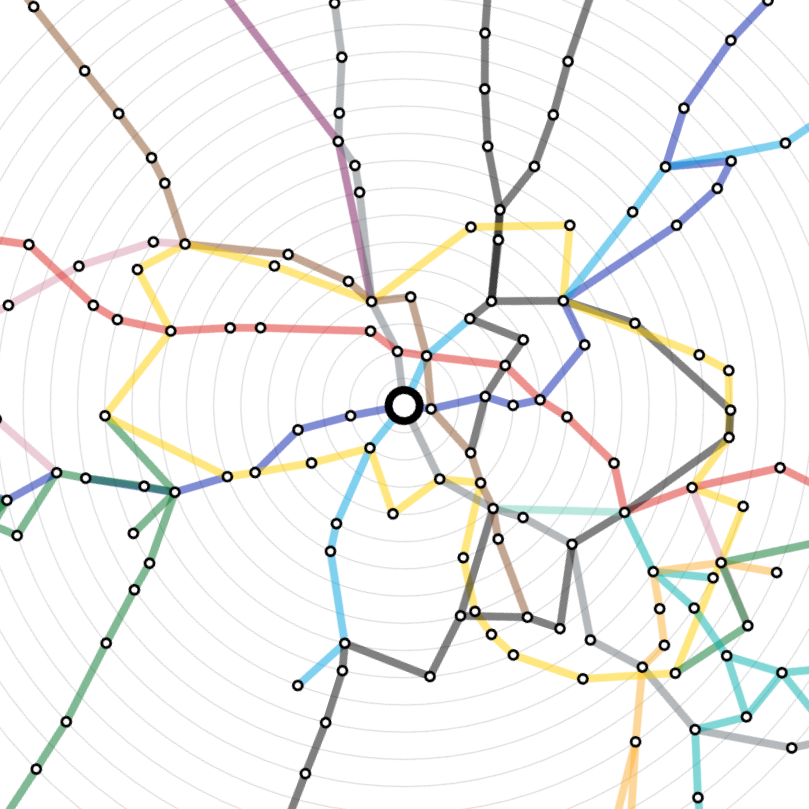}
    \caption{The London Underground map with a focus on Green Park, created using the method described in Section~\ref{focus}. The distances are based on travel time rather than real world distance.
    Data from~\cite{tube}.}
    \label{radial}
\end{figure}

It is often the case that a user will want to examine specific vertices in a graph, especially in an interactive setting. It is therefore important to be able to emphasize distances involving certain vertices.
Brandes and Pich~\cite{brandes2011more} presented a general method of doing this in the context of majorization, by interpolating between two stress summations representing general and constrained weights separately.

For SGD, emphasizing specific distances is as simple as weighting the corresponding constraints more heavily.
For example to focus on vertex $3$, we simply set the relevant weights to infinity
\begin{equation}
w_{ij} \leftarrow \infty \quad \text{if}\ i=3\ \text{or}\ j=3.
\end{equation}
This causes only the remaining constraints to decay, but the system still converges in this case as there are no conflicts between the ones emphasized.
Setting weights to infinity when using majorization results in the algorithm becoming instantly very stuck, which is why the more complicated interpolation~\cite{brandes2011more} is necessary.

These emphasized distances can also be modified from their
graph-theoretic
values if a specific separation is desired, for example to constrain in a circle using the distances introduced by Dwyer~\cite{dwyer2009scalable}. Additional constraints such as directed edges or non-overlap boundaries~\cite{dwyer2009scalable} can also be added just as easily by changing the objective function as desired.

\subsection{Color and Interactivity}

\begin{figure}
    \centering
    \includegraphics[height=.85\linewidth]{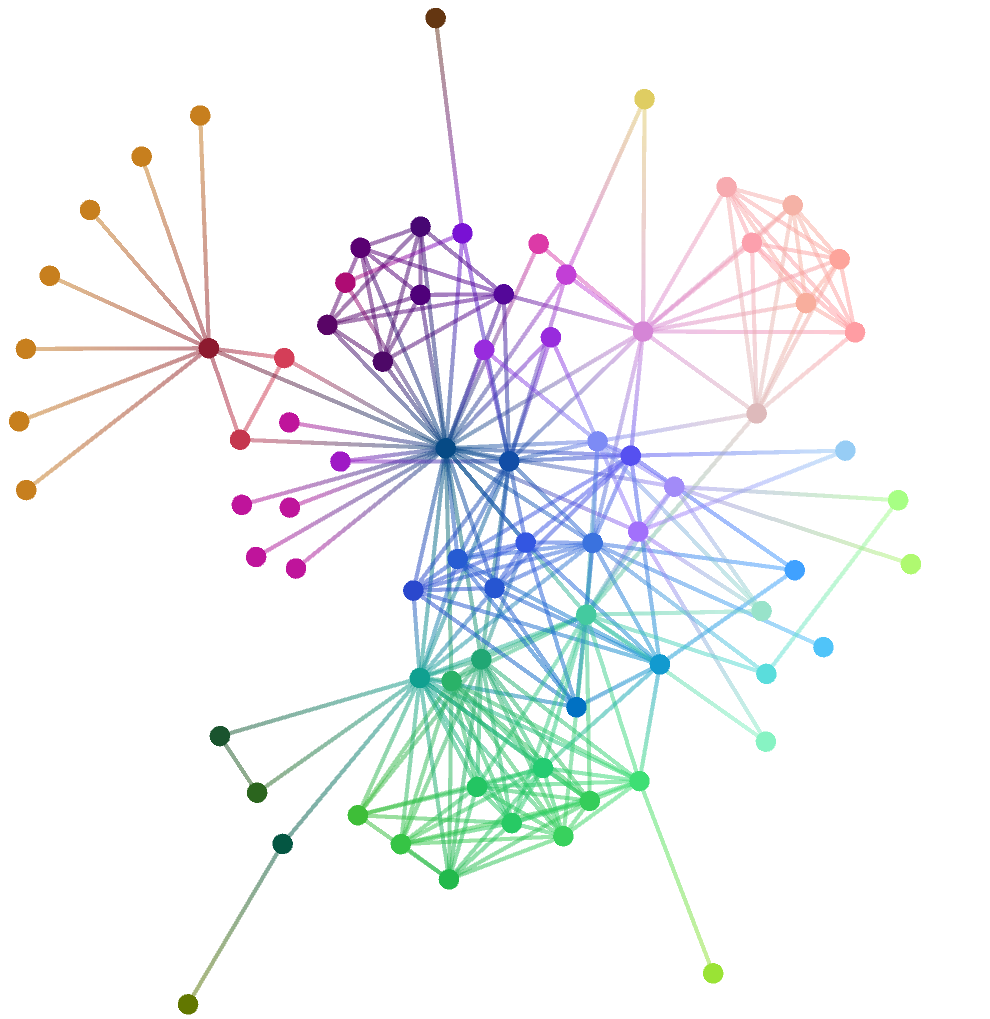}
    \caption{
    Co-appearances of characters in Les Miserables by Victor Hugo.
	Groups of similarly colored vertices indicate clustering based on Jaccard similarity.
}
    \label{jaccard}
\end{figure}

Highly connected and small-world graphs such as social networks can often produce dense, entangled layouts colloquially termed `hairballs'. In this case, it is often useful to try to uncover some other form of information, such as revealing clusters of similar vertices. Since color is simply a
linear
mix of red, green, and blue (RGB), it can be used as a three-dimensional space in which Euclidean distances can be embedded, where each color corresponds to a separate axis.
Figure~\ref{jaccard} shows an example of vertices colored by their Jaccard similarity index, defined as
\begin{equation}
d_{ij} = 1 - \frac{|N(i) \cap N(j)|}{|N(i) \cup N(j)|}
\end{equation}
where $N(i)$ are the neighbors of vertex $i$. Since $d_{ij}$ is bounded between 0 and 1, embedded distances fit perfectly within the similarly bounded axes of color. This means that vertices not only have coordinates within normal Euclidean space, but also within RGB space.

This process can help to reveal groupings, but can also produce ambiguity when applied to larger graphs due to the lack of distinct color combinations, again a problem caused by a lack of output dimensions. One possibility in this case would be to use an interactive form of visualization in which the user selects a smaller group of vertices at a time, and the algorithm embeds only their selection in an RGB space, by considering dissimilarities only between selected vertices.

Interaction could also allow the user to manually adjust the step size $\eta$ from Equation~(\ref{SGD step}), allowing them to `shake' the graph out of local minima themselves. The step size annealing from Section~\ref{cooling} is the most ad hoc and data-dependent component of SGD so handing control over to the user can be useful, especially in dynamical situations where the structure of the graph changes with time.
Additionally, if frame rate becomes an issue in an interactive setting, the application does not have to wait until the end of an entire iteration before rendering an updated layout because vertices are continually being moved, keeping the user interface smooth and responsive.

\subsection{Large Graphs}
\label{sparse}
To understand how many layout algorithms tackle scaling to larger graphs, it is convenient to rewrite Equation~(\ref{stress}) by splitting the summation into two parts: paths that traverse one edge, and paths that traverse multiple.
With $\sigma_{ij} = (||\mathbf{X}_i - \mathbf{X}_j|| - d_{ij})^2$
this is
\begin{equation}
\label{split stress}
\text{stress}(\mathbf{X}) = \sum_{\{i,j\}\in E}w_{ij}\sigma_{ij}\\
+ \sum_{\{i,j\}\notin E}w_{ij}\sigma_{ij}
\end{equation}
where $E$ is the set of edges in the graph.
Just considering the preprocessing stage for now, it is clear that we can easily compute $d$ and $w$ for the first half of the summation directly from the graph. Real-world graphs are also usually sparse, so for a graph with $n$ vertices and $m$ edges, $m \ll n^2$ making the space required to store these values tolerable. However the second half is not so easy---an all-pairs shortest paths (APSP) calculation takes $O(m + n)$ time per vertex for an unweighted graph with a breadth-first search, or $O(m + n\log n)$ for a weighted graph using Dijkstra's algorithm~\cite{cormen2009introduction}. Combined with requiring $O(n^2)$ space to store all the values of $d_{ij}$, this makes the preprocessing stage alone intractable for large graphs.

The second stage is iteration, where the layout is gradually improved towards a good minimum. Again, computing the first summation is tolerable, but the number of longer distance contributions quickly grows out of control. Many notable attempts have been made at tackling this second half. A common approach is to ignore $d_{ij}$, and to approximate the summation as an $n$-body repulsion problem, which can be efficiently well approximated using $k$-d trees \cite{barnes1986hierarchical}. Hu~\cite{hu2005efficient} and independently Hachul and J\"unger~\cite{hachul2004drawing} used this in the context of the force-directed model of Fruchterman and Reingold~\cite{fruchterman1991graph}, along with a multilevel coarsening scheme to avoid local minima. Gansner et al.~\cite{gansner2013maxent} use it with majorization by summing over $-\alpha \log||X_i - X_j||$ instead. Brandes and Pich~\cite{brandes2006eigensolver} even ignore the second half completely and capture the long-range structure by first initializing with a fast approximation to \emph{classical scaling}~\cite{brandes2008experimental}, which minimizes the inner product rather than Euclidean distance.

There are a couple of issues with this idea, one being that treating all long-range forces equally is unfaithful to
graph-theoretic distances,
and another being that the relative strength of these forces depends on an extra parameter that can strongly affect the final layout of the graph~\cite{hu2005efficient}.
Keeping these dependent on their graph-theoretic distance sidesteps both of these issues, but brings back the problem of computing and storing shortest paths. One approach to maintaining this dependence comes from Khoury et al.~\cite{khoury2012drawing}, who use a low-rank approximation of the distance matrix based on its singular value decomposition. This can work extremely well,
but still requires APSP unless $w_{ij} = d_{ij}^{-1}$.

\begin{figure}
  \removelatexerror

  \begin{algorithm}[H]
  \SetKwProg{SGD}{SparseSGD}{}{} \SetKwInOut{Input}{inputs}
  \SetKwInOut{Output}{output}
  \SetKwFor{ForEach}{foreach}{:}{}
  \SetKwFor{For}{for}{:}{}
  \SetKwIF{If}{ElseIf}{Else}{if}{:}{elif}{else}{}

  \SGD{\emph{\textbf{(}}$G,h$\emph{\textbf{):}}}{
      \Input{graph $G=(V,E)$, number of pivots $h$}
      \Output{$k$-dimensional layout
      $\mathbf{X}$ with $n$ vertices}

      $P \leftarrow MaxMinRandomSP(G,h)$
      \label{code:maxminrandom}

      $d_{\{p,i\}} \leftarrow SparseShortestPaths(G,P)$
      \label{code:sparseshortestpaths}
      
      $w_{\{p,i\}}' \leftarrow 0$

      \ForEach{$\{p,i:p \notin N(i)\} \in (P \times V)$}{
          $s \leftarrow |\{j\in R(p): d_{pj} \leq d_{pi}/2\}|$


          $w_{ip}' \leftarrow s \, w_{ip}$
          \label{code:s}
      }

      \ForEach{$\{i,j\} \in E$}{
          $w_{ij}' \leftarrow w_{ji}' \leftarrow w_{ij}$
      }

      $\mathbf{X} \leftarrow RandomMatrix(n,k)$


      \For{$\eta$ in annealing schedule}{
          \ForEach{$\{i,j\} \in E \cup (V \times P)$ in random order}{

              $\mu_i \leftarrow Min(w_{ij}' \eta \,,\ 1)$

              $\mu_j \leftarrow Min(w_{ji}' \eta \,,\ 1)$

              $\mathbf{r} \leftarrow \frac{||\mathbf{X}_i - \mathbf{X}_j|| - d_{ij}}{2}\frac{\mathbf{X}_i - \mathbf{X}_j}{||\mathbf{X}_i - \mathbf{X}_j||}$

              $\mathbf{X}_i \leftarrow \mathbf{X}_i - \mu_i \mathbf{r}$

              $\mathbf{X}_j \leftarrow \mathbf{X}_j + \mu_j \mathbf{r}$
          }
      }
  }

  \caption{Sparse SGD}
  \label{sparse pseudo}
  \end{algorithm}

  \caption{Pseudocode for performing SGD on the sparse stress approximation described in Section~\ref{sparseexplanation}.
  Note that all $R(p)$ and $w_{ip}'$ can be constructed over the course of shortest path calculations without increasing the asymptotic complexity~\cite{ortmann2016sparse}.
  \label{algorithm2}
  }
\end{figure}

\subsubsection{Sparse Approximation}
\label{sparseexplanation}

The approach we use is that of Ortmann et al.~\cite{ortmann2016sparse}, who pick a set of pivots whose shortest paths are used as an approximation for the shortest paths of vertices close to them. Since this approach actually reduces the number of terms in the summation, using it in the context of
SGD
also reduces the amount of work per iteration.

To approximate the full model well it is important to choose pivots that are well distributed over the graph, and in their original paper Ortmann et al.~\cite{ortmann2016sparse} present an experimental evaluation of various methods for doing so. Our implementation uses \textit{max/min random sp} to select pivots.
Non-random \textit{max/min sp} starts by picking one or more pivots and computing their shortest paths to all other vertices, with subsequent pivots chosen by picking the vertex with the maximum shortest path to any pivot chosen so far~\cite{de2004sparse}. The random extension instead samples for subsequent pivots with a probability proportional to this shortest path to any pivot, rather than simply always picking the maximum.

\begin{figure*}
    \centering
    \includegraphics[width=\textwidth]{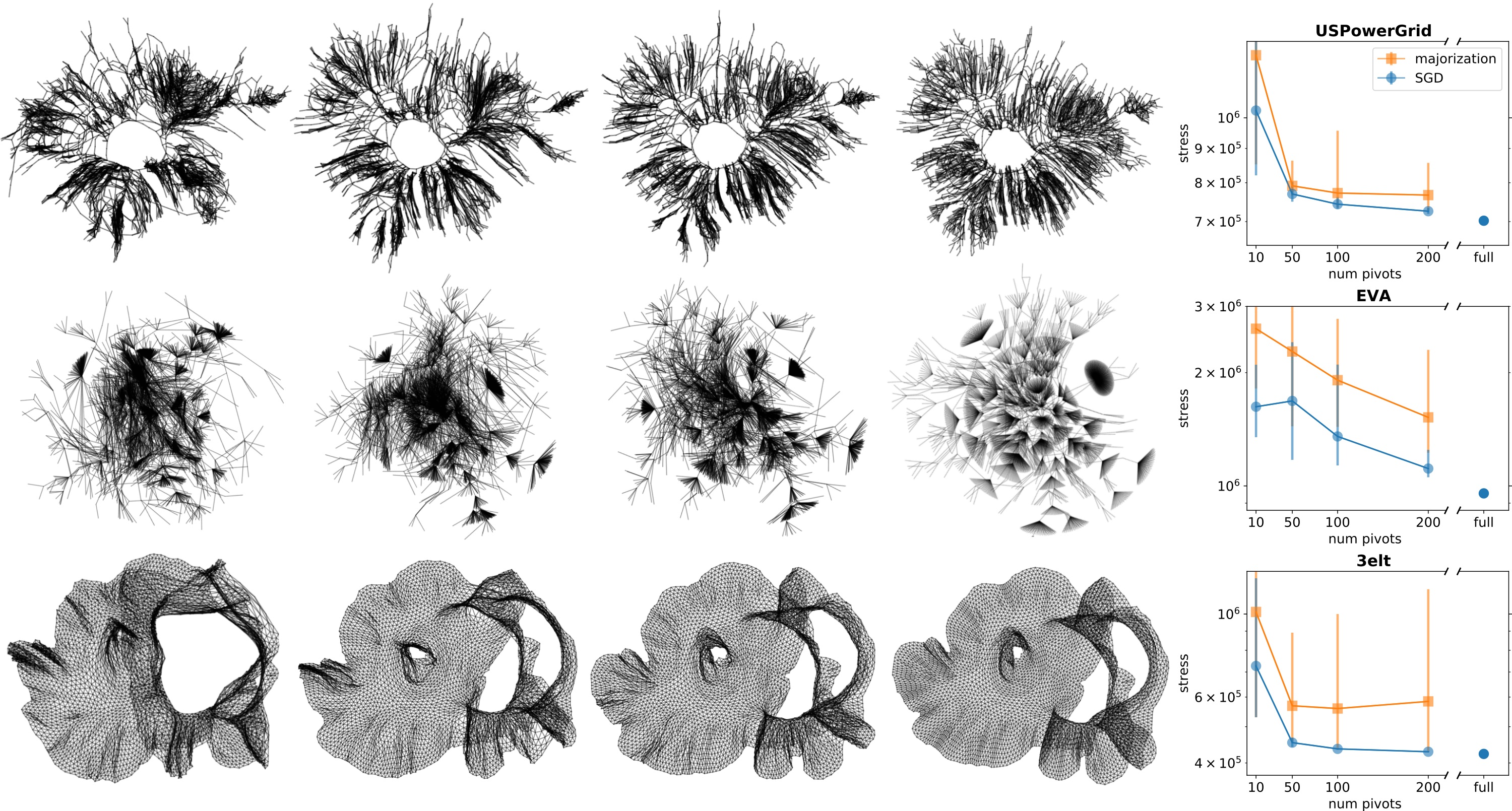}
    \caption{Examples of sparse relaxation on the graphs \texttt{USPowerGrid}, \texttt{EVA}, and \texttt{3elt}. From left to right: layouts from 10 pivots, 50, 200, full stress, and plots showing stress over 25 runs for each number of pivots.
    We use the 15 iteration annealing schedule from Section~\ref{fixediterations} for SGD, and allow majorization to run for 100 iterations.
    Layouts shown contain the minimum stress achieved for that number of pivots.
    \texttt{EVA} is the least well approximated by the sparse model, likely due to its low diameter and high degree distribution.
    The mesh-like \texttt{3elt} is very well approximated by the model itself, but the performance of majorization declines as the number of pivots exceeds $\sim$100, likely due to the increased number of terms in the summation that introduce more local minima that majorization struggles to jump over.
    Note that we did not use PivotMDS~\cite{brandes2006eigensolver} to initialize as done by Ortmann et al.~\cite{ortmann2016sparse}, and instead initialize randomly within a 1$\times$1 square as before.
    }
    \label{pivots}
\end{figure*}

\begin{figure*}
    \centering
    \includegraphics[width=\textwidth]{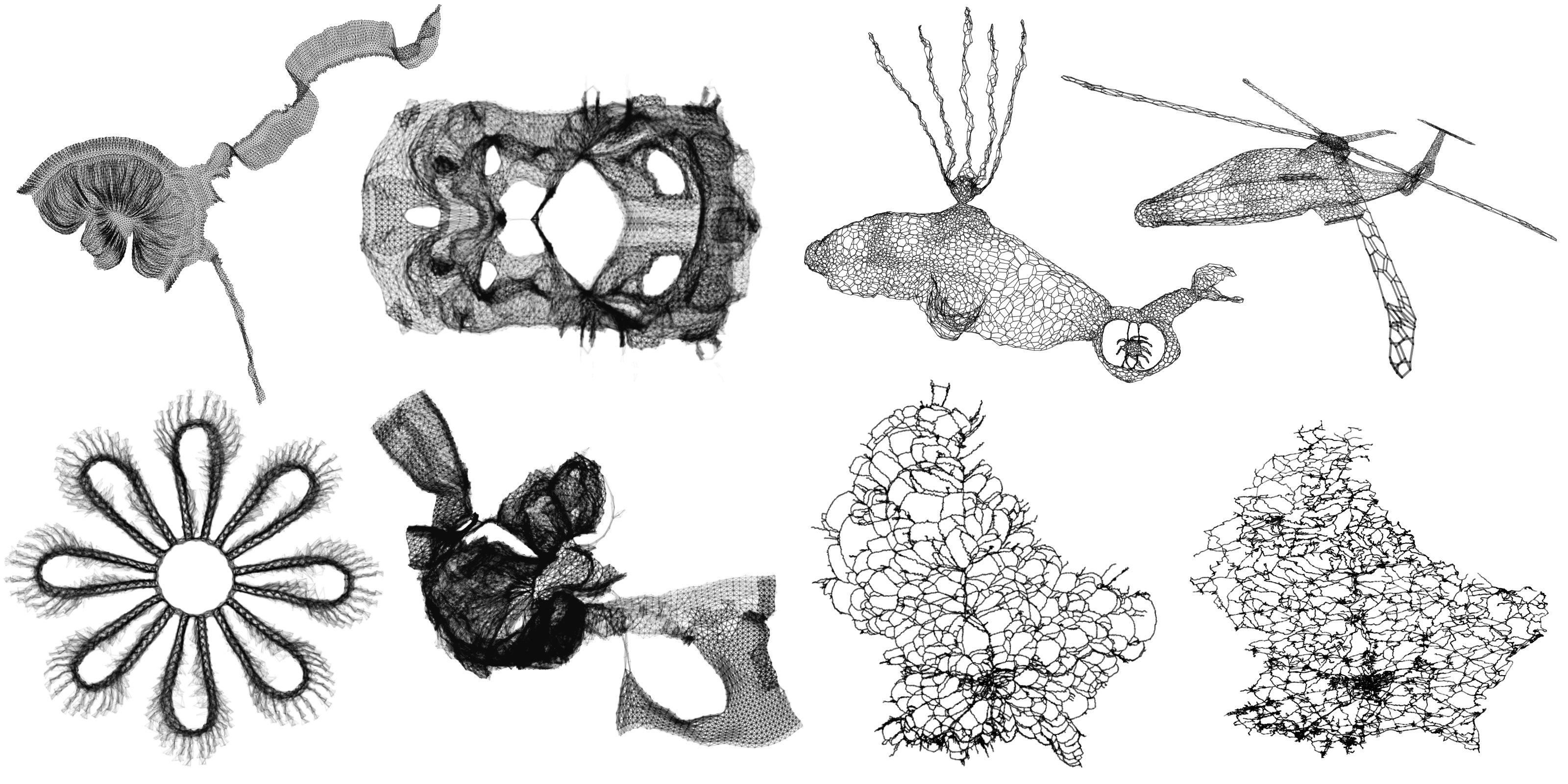}
    \caption{Some larger graphs, each approximated with 200 pivots, also using the 15 iteration schedule from Section~\ref{fixediterations}. From left to right, top row then bottom: \texttt{pesa} (11,738 vertices), \texttt{bcsstk31} (35,588), \texttt{commanche\_dual} (7,920) and its original layout, \texttt{finance256} (37,376), \texttt{bcsstk32} (44,609), \texttt{luxembourg\_osm} (114,599) and its original layout. The right-hand graphs have path lengths weighted according to distances calculated from the original layouts.
    }
    \label{showcase}
\end{figure*}

These pivots $p\in P$ are then each assigned a region $R(p)$, which is the set of vertices closer to that pivot than any other. The relevant weights $w_{ip}$ are then adapted depending on the composition of the region, resulting in a new decomposed second half of the summation
\begin{equation}
\label{pivot stress}
\begin{split}
\text{stress}(\mathbf{X}) = \sum_{\{i,j\}\in E}w_{ij}\sigma_{ij}
+ \sum_{i\in V}\sum_{p\in P\setminus N(i)}w_{ip}'\sigma_{ij}
\end{split}
\end{equation}
where $N(i)$ are the neighbors of $i$ to prevent overlap with any edges in the first summation.
The adapted weight $w_{ip}'$ is then set to $s_{ip} w_{ip}$, where $s_{ip}$ is the number of vertices in $R(p)$ at least as close to $p$ as to $i$:
\begin{equation}
\label{s}
s_{ip}=|\{j\in R(p): d_{jp} \leq d_{ip}/2\}|
\end{equation}
The reason the weight on vertex $i$ is increased like this is because its contribution acts as an approximation for the stress to all vertices in $R(p)$, and (\ref{s}) is required to prevent the weight on closer vertices from being overestimated.
It is important to note that if both vertices $p$ and $q$ are pivots then $w_{pq}'$ may not equal $w_{qp}'$ and if only $p$ is a pivot then $w_{pq}'=0$ as $q$ should not contribute to the position of $p$.
Resulting layouts are presented in Figures~\ref{pivots} and~\ref{showcase}, and pseudocode can be seen in Algorithm~\ref{sparse pseudo} (Figure~\ref{algorithm2}).

\section{Discussion}

One of the major reasons why previous force-directed algorithms, such as in~\cite{fruchterman1991graph,kamada1989algorithm, dwyer2009scalable},
have become popular is how simple and intuitive the concept is. The idea of a physical system pushing and pulling vertices, modeled as sets of springs and electrical forces, makes them easy to understand and quick to implement for practical use.

The geometric interpretation of the SGD algorithm we have presented
shares these qualities, as the concept of moving pairs of vertices one by one towards an ideal distance is just as simple.
In fact the stress formulation~(\ref{stress}) is commonly known as the spring model~\cite{kamada1989algorithm,hu2005efficient}, and the physical analogy of decompressing one spring at a time very naturally fits this intuition.
The implementation also requires no equation solver, and there is no need to consider smart initialization, which can often be just as complex a task \cite{brandes2008experimental}.
Considering only a single pair of vertices at a time also makes further constrained layouts easy to implement, and allows an appropriate sparse approximation to grant scalability up to large graphs.

But perhaps the most important benefit of SGD is its consistency regardless of initialization, despite being non-deterministic due to the shuffling of the order of terms. By contrast, the plots in Section~\ref{results} clearly show how vastly the results from majorization can differ depending on initialization, especially when restricted to a limited number of iterations. This reliability of SGD can be crucial for real-time applications with fixed limits on computation time, such as within an interactive visualization.

However there are still situations where SGD can struggle with local minima, such as \texttt{dwt\_2680} which is susceptible to twisting in the middle. This can be seen in Figure~\ref{appendix} where we purposefully included a twisted layout to illustrate this pitfall.
A potential solution to this is overshooting, or in other words allowing values of $0 < \mu < 2$ in Equation~(\ref{mu}). This greatly reduces the chance of a twist, but results in poorer local minima in most other cases and can also bring back the problem of divergence, so is a potential avenue for future work,
perhaps to be used in conjunction with an adaptive annealing schedule to further optimize performance depending on the input data.



\subsection{Conclusion}
In this paper we have presented a modified version of stochastic gradient descent (SGD) to minimize stress as defined by Equation~(\ref{stress}). An investigation comparing the method to majorization shows consistently faster convergence to lower stress levels, and the fact that only a single pair of vertices is considered at a time makes it well suited for variants such as constrained layout or the pivot-based approximation of Ortmann et al.~\cite{ortmann2016sparse}.
This improved performance---combined with a simplicity that forgoes an equation solver or smart initialization---makes
SGD a strong candidate for general graph layout applications.

Code used for timing experiments, along with some example Jupyter notebooks, is open source and available at \url{www.github.com/jxz12/s_gd2}.

\appendices

\ifCLASSOPTIONcompsoc
  \section*{Acknowledgments}
\else
  \section*{Acknowledgment}
\fi

We thank Tim Davis and Yifan Hu for maintaining the SuiteSparse Matrix Collection \cite{davis2011university}, where most of the graph data used in this paper was obtained.
We are also grateful to the anonymous reviewers whose comments helped us to improve the paper.

\ifCLASSOPTIONcaptionsoff
  \newpage
\fi



\bibliographystyle{IEEEtran}
\input{main.bbl}

%

%

\begin{IEEEbiography}
[{\includegraphics[width=1in,height=1.25in,clip,keepaspectratio]{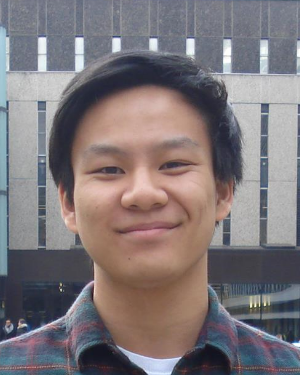}}]
{Jonathan X. Zheng} (corresponding author)
is a PhD student in the Department of Electrical and Electronic Engineering, Imperial College London. His interests lie in the visualization of complex networks, along with its application to serious games to crowdsource research through public engagement. Zheng received his MEng in Electronic Information Engineering from Imperial College London.
\end{IEEEbiography}


\begin{IEEEbiography}
[{\includegraphics[width=1in,height=1.25in,clip,keepaspectratio]{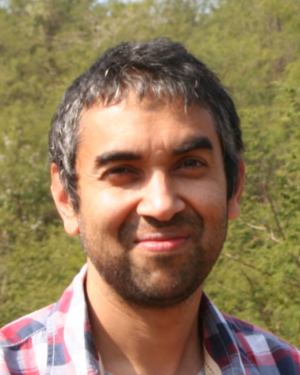}}]
{Samraat Pawar}
is a Senior Lecturer in the Department of Life Sciences, Imperial College London. His main research interests are in computational and theoretical biology, with particular focus on the dynamics of complex ecosystems and underlying interaction networks. Pawar received his PhD in Biology from the University of Texas at Austin.
\end{IEEEbiography}

\begin{IEEEbiography}
[{\includegraphics[width=1in,height=1.25in,clip,keepaspectratio]{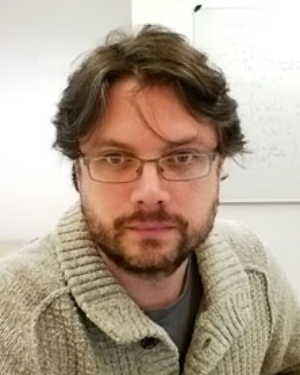}}]
{Dan F. M. Goodman}
is a Lecturer in the Department of Electrical and Electronic Engineering, Imperial College London. His main research interests are in computational neuroscience, and he is also interested in studying dynamically evolving network structures such as ecosystems. Goodman received his PhD in Mathematics from the University of Warwick.
\end{IEEEbiography}




\end{document}